\documentclass[12pt]{article}
\usepackage{epsfig}

\renewcommand{\thefootnote}{\fnsymbol{footnote}}
\begin{document}

\title{\ \\*[10pt]
{\bf \Large 
Patterns of supersymmetry breaking in 
moduli-mixing racetrack model\footnote{
Talk given at 
the 14th International Conference on Supersymmetry 
and the Unification of Fundamental Interactions (SUSY06), 
UC Irvine, California, USA, 12-17 June 2006 
and 
the International Workshop on Supersymmetry, 
Electroweak Symmetry Breaking and Particle Cosmology 
(Summer Institute 2006), 
APCTP, Pohang, Korea, 21-30 August 2006.}}\\*[20pt]
}

\author{Hiroyuki~Abe\footnote{
E-mail address: abe@yukawa.kyoto-u.ac.jp} \\*[20pt]
{\it \normalsize 
Yukawa Institute for Theoretical Physics, 
Kyoto University, Kyoto 606-8502, Japan} \\*[50pt]}

\date{
\centerline{\small \bf Abstract}
\begin{minipage}{0.9\linewidth}
\medskip 
\medskip 
\small
We show some structures of moduli stabilization and supersymmetry 
breaking caused by gaugino condensations with the gauge couplings 
depending on two moduli which often appear in the four-dimensional 
effective theories of superstring compactifications. 
\end{minipage}
}

\begin{titlepage}
\maketitle
\thispagestyle{empty}
\end{titlepage}

\renewcommand{\thefootnote}{\arabic{footnote}}
\setcounter{footnote}{0}

\section{Introduction}

Superstring/M-theory is a candidate of unified description 
for elementally particles and their interactions including 
gravity. Soft supersymmetry (SUSY) breaking parameters 
in the four-dimensional (4D) effective theory as well as 
the 4D Planck scale $M_{Pl}$, gauge couplings and Yukawa 
couplings are given by the vacuum expectation values (VEVs) 
of modulus superfields which determine the size and shape 
of extra dimensions. Then the moduli stabilization and its 
effect on SUSY breaking is quite relevant to the particle 
phenomenology and cosmology. Here we show some structures 
of moduli stabilization and SUSY breaking in a so-called 
racetrack model with double gaugino condensations where 
gauge couplings are given by more than one modulus 
field~\cite{Abe:2005rx,Abe:2005pi,Abe:2006xi} 
(two moduli in practice). 

\section{Moduli-mixing racetrack model}

The 4D effective supergravity (within type IIB $O3/O7$ 
framework for concreteness if necessary) is given by 
the K\"ahler potential and superpotential 
\begin{eqnarray}
K=-n_T \ln (T+\bar{T})-n_S \ln (S+\bar{S}), \quad 
W=Ae^{-af_a}-Be^{-bf_b}, 
\label{eq:4dsugra}
\end{eqnarray}
where 
$f_{a,b}=m_{a,b}S+w_{a,b}T$. 
The superfields $S$ and $T$ represents the dilaton and the 
K\"ahler (size) modulus respectively. The $n_S$ and $n_T$ 
are some model dependent numbers given typically by 
$(n_S,n_T)=(1,3)$, and $m_{a,b}$, $w_{a,b}$ are respectively 
the magnetic flux and the winding number of the 
$D$-brane~\cite{Lust:2004cx} where 
the gaugino condensation occurs, which generates the 
nonperturbative superpotential. For example, gaugino 
condensation on the $D3$-, $D7$-, magnetized $D7$- and 
magnetized $D9$-brane ($D3$ is assumed to be located at 
some singular point where the extended SUSY is reduced) 
yields the nonperturbative superpotential (\ref{eq:4dsugra}) 
with $f=S$, $T$, $|m|S+|w|T$ and $mS-wT$, respectively. 

Before arriving at the above effective theory (\ref{eq:4dsugra}) 
for the light moduli $S$ and $T$, we have assumed that the existence 
of three-form flux, $G_3=F_3-2 \pi iS H_3$, in ten-dimensions stabilizes 
the complex structure (shape) moduli $U$ as $\langle U \rangle \sim 1$ 
at around the Planck scale in a supersymmetric way, $D_U W_{\rm flux}=0$, 
through the superpotential~\cite{Gukov:1999ya} 
\begin{eqnarray}
W_{\rm flux} &=& \int_{CY_3} G_3 \wedge \Omega 
\ = \ f^{RR}(U)+Sf^{NS}(U), 
\nonumber
\end{eqnarray}
where $\Omega$ is a holomorphic three-form of the 
Calabi-Yau (CY) three-fold, and $f^{RR,NS}(U)$ are 
some functions of $U$ determined by the flux. 
Note also that due to the flux there is a significantly warped 
region in the CY space~\cite{Giddings:2001yu}. 

\subsection{Single light modulus}

If the three-form flux induces a SUSY mass like 
$W_{\rm flux} \sim SU$, 
that is $n_{RR}=n_{NS}=1$ in 
\begin{eqnarray}
f^{RR,NS}(U) &=& 
(U-\langle U \rangle)^{n_{RR,NS}} \tilde{f}^{RR,NS}(U), 
\qquad 
\tilde{f}^{RR,NS}(\langle U \rangle) \ \ne \ 0, 
\label{eq:fluxsp}
\end{eqnarray}
the dilaton is also stabilized 
$\langle S \rangle 
=-\tilde{f}^{RR}( \langle U \rangle)
/\tilde{f}^{NS}( \langle U \rangle) \sim 1$ 
as well as $U$, via the global SUSY vacuum conditions 
$\partial_{U,S}W_{\rm flux}=W_{\rm flux}=0$~\cite{Abe:2006xi}. 
In this case we replace $S$ in Eq.~(\ref{eq:4dsugra}) by its VEV, 
$\langle S \rangle$. Then the effective superpotential becomes 
\begin{eqnarray}
W &=& A'e^{-aw_aT}-B'e^{-bw_bT},
\label{eq:single}
\end{eqnarray}
which is in the same form as the racetrack model with single 
modulus, but the coefficients are exponentially suppressed 
or enhanced~\cite{Abe:2005rx}, 
$A'=Ae^{-am_a \langle S \rangle}$, 
$B'=Be^{-bm_b \langle S \rangle}$, 
where $a=8 \pi^2/N_a$ and $b=8\pi^2/N_b$ 
for $SU(N_{a,b})$ gaugino condensation. 

The minimum of the scalar potential induced by the above 
K\"ahler and superpotential corresponds to a SUSY AdS$_4$ 
local minimum with negative vacuum energy and 
$a \langle  \textrm{Re}\,T \rangle
=at_{SUSY} \sim \ln (M_{Pl}/m_{3/2})$, 
where $m_{3/2} \approx 10^{-14}M_{Pl}$ is the gravitino mass. 
To be phenomenologically viable, we uplift the vacuum energy 
by introducing anti $D3$-branes at the top of warped region 
in the CY space~\cite{Kachru:2003aw}. 
Then the SUSY is broken due to the slight 
shift $\delta T = (t_{SUSY}-\langle T \rangle) \ll t_{SUSY}$ 
caused by an additional potential energy of 
$\overline{D3}$s~\cite{Choi:2005ge}. 

In this case, the ratio between the VEV of auxiliary component 
in the chiral compensator $C$, $F^C \sim m_{3/2}$, and one 
in $T$, $F^T$, is given by~\cite{Abe:2005rx} ($w_a=0$ for simplicity) 
$$
\displaystyle 
\alpha \ = \ 
\frac{F^C}{\ln(M_{Pl}/m_{3/2})} \, 
\frac{T+\bar{T}}{F^T} 
\ \simeq \ 
\frac{1}{1+m_b \langle S \rangle /w_b t_{SUSY}}. 
$$ 
This corresponds to the ratio between the so-called 
anomaly mediation and the modulus mediation for the 
visible sector SUSY breaking. It was shown in Ref.~\cite{Choi:2005uz} 
that the sparticle masses are unified at the scale given by 
$\Lambda_m = e^{-2 \pi \alpha} \Lambda$ where $\Lambda$ 
is a messenger scale of the modulus mediated contribution. 
This $\Lambda_m$ is called a mirage messenger scale, and  
if $\alpha \sim {\cal O}(1)$, the $\Lambda_m$ is quite 
lower than the $\Lambda$. 
Here we find that $\alpha$ varies in a wide range with 
various magnetic flux $m_b$, compared to $\alpha \sim 1$ 
without the magnetic flux in the original analysis~\cite{Choi:2005ge}. 

Another implication is that a runaway structure of the 
usual racetrack model can be avoided for the negative value 
of $w_{a,b}$ in Eq.~(\ref{eq:single}) that may be realized, e.g., 
by the gaugino condensation on the magnetized $D9$-brane. 
In this case, the scalar potential is lifted in the region 
${\rm Re}\,T \gg t_{SUSY}$. This situation can be a solution 
to the destabilization/overshooting problem with a finite 
temperature effect and with an arbitrary initial condition for $T$.

\subsection{Two light moduli}

\begin{figure}
\epsfig{figure=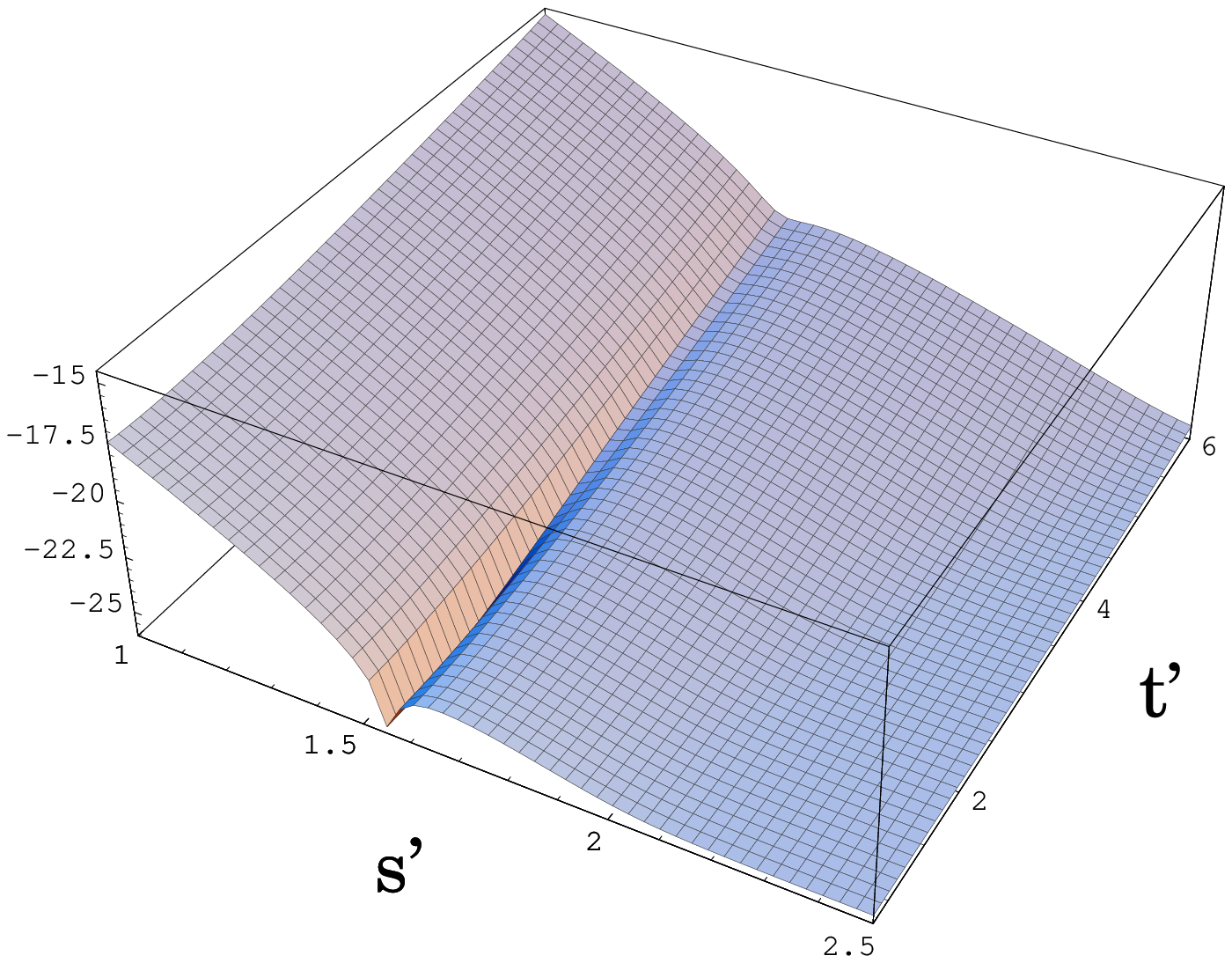,width=.48\linewidth}
\hspace{.02\linewidth}
\epsfig{figure=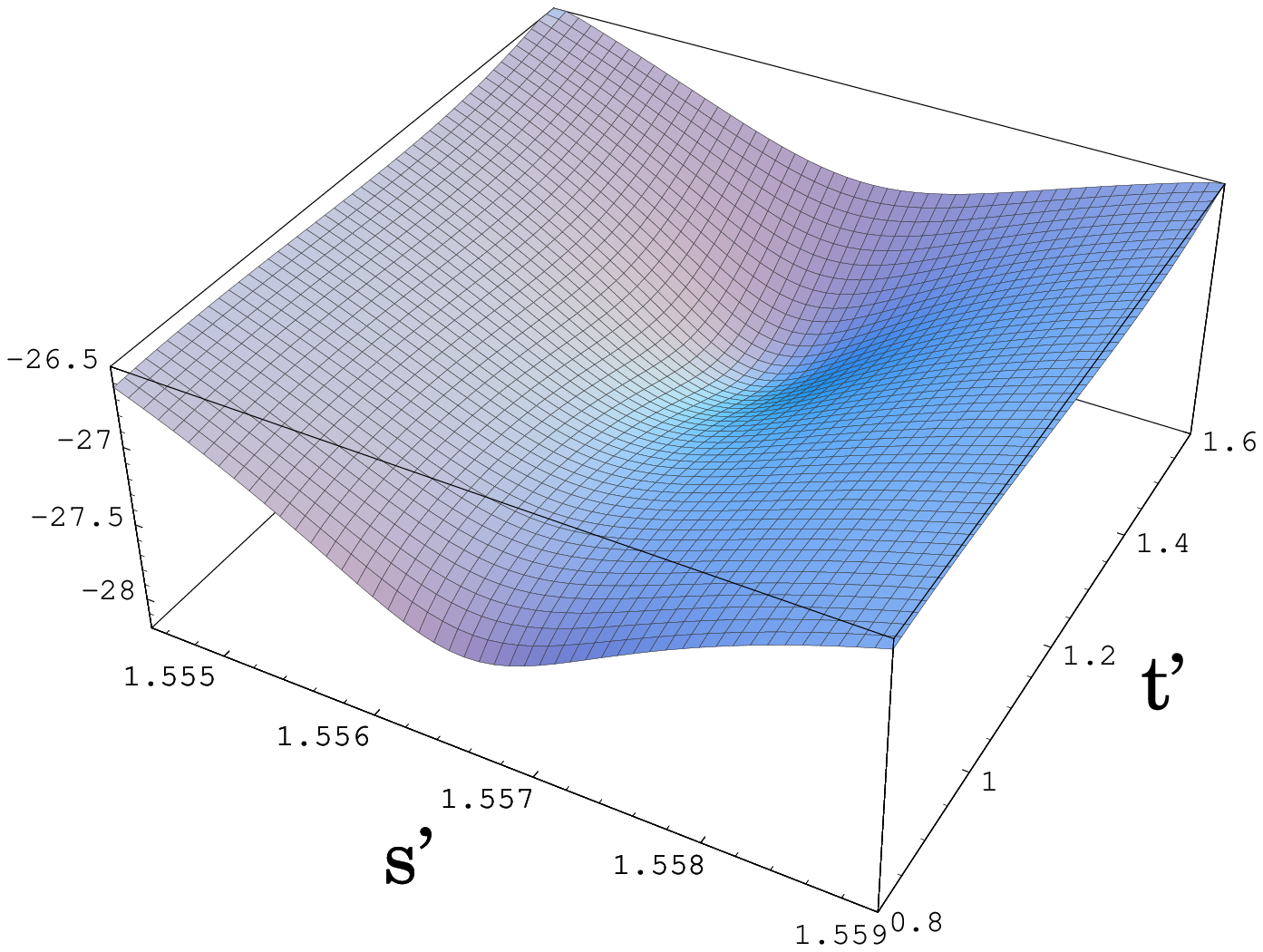,width=.48\linewidth}
\caption{The logarithm of the scalar potential for typical 
values of parameters. The primed $s$ and $t$ mean that the 
$(s,t)$-plane is rotated around the origin in such a way 
that the `racetrack' becomes perpendicular to the $s'$-axis. 
Note that the minimum of the potential is shifted 
to be positive inside the logarithm.}
\label{fig:sbmin}
\end{figure}

When the three-form flux $G_3$ does not contain 
the SUSY mass term $W_{\rm flux} \sim SU$, that is 
$n^{RR}=n^{NS}=2$ in Eq.~(\ref{eq:fluxsp}), 
the dilaton remains as a light modulus as well as $T$ 
in Eq.~(\ref{eq:4dsugra})~\cite{Abe:2006xi}. 
A careful analysis for Eq.~(\ref{eq:4dsugra}) 
shows~\cite{Abe:2005pi} that a SUSY AdS$_4$ 
stationary point is given by 
\begin{eqnarray}
s_{SUSY} &\sim& \frac{1}{b-a}\ln\frac{bB}{aA}, \qquad 
t_{SUSY} \ \sim \ \frac{n_T(b-a)}{2ab(\alpha+\beta)}, 
\nonumber
\end{eqnarray}
which is valid within a parameter region satisfying 
$|s_{SUSY}| \gg n_S/2a,\, n_S/2b$, where 
the paremeters $a$, $b$, $w_a$ and $w_b$ in 
Eq.~(\ref{eq:4dsugra}) are replaced by
$a/m_a$, $b/m_b$, $-m_a \alpha$ and $m_b \beta$, 
and the small letters $s$ and $t$ stand for 
${\rm Re}\,S$ and ${\rm Re}\,T$, respectively. 
In order this SUSY point to reside in a perturbative region 
$s_{SUSY},t_{SUSY}>1$ with $a,b \gg 1$ and $A,B \sim 1$, 
we need a fine-tuning like $b-a\sim{\cal O}(1)$, 
$\alpha,\beta \sim {\cal O}(a^{-2}),{\cal O}(b^{-2})$. 
Moreover, this SUSY point is actually a saddle point 
located in a sharp `racetrack' of the scalar potential 
which is implied by the hierarchical mass square eigenvalues 
$(m_\bot^2,\,m_\parallel^2)$ at this point 
with $m_\bot^2>0$, $m_\parallel^2<0$ and 
$-m_\bot^2/m_\parallel^2 \sim 
\frac{128}{n_S^2} \left( \ln \frac{bB}{aA} \right)^4 
\frac{(ab)^2}{(b-a)^4} \ \gg \ 1$. 

A SUSY breaking AdS$_4$ local minimum $(s_{SB}, t_{SB})$ 
exists along the same `racetrack' and given by 
$s_{SB}=s_{SUSY}(1+\delta_{SB}^s)$ and 
$t_{SB}=t_{SUSY}(1+\delta_{SB}^t)$ where 
\begin{eqnarray}
\delta_{SB}^t &\sim& \frac{\sqrt{n_T+1}-1}{n_T}, \qquad 
\delta_{SB}^s \ \sim \ 
-\frac{a\alpha+b\beta}{\ln (bB/aA)}\,\delta_{SB}^t. 
\nonumber
\end{eqnarray}
By uplifting this local minimum with some lifting potential, 
we can obtain a Minkowski minimum but with modulus-dominated 
SUSY breaking $\alpha \ll 1$. 
This is because SUSY is broken before uplifting 
unlike the previous single modulus case. 

Fig.~\ref{fig:sbmin} shows a numerical plot 
of the scalar potential in $(s,t)$-plane. 
We can see the sharp `racetrack' structure, 
and the two stationary points along this. 

\section{Conclusions}

We have shown a vacuum structure of the racetrack model 
with double gaugino condensations where the gauge couplings 
depend on two typical moduli $S$ and $T$ appearing in the 
superstring compactifications. If the flux induced superpotential 
$W_{\rm flux}$ stabilizes one of them, $S$, at a high scale, 
the nonperturbative superpotential in Eq.~(\ref{eq:single}) can 
have large suppression/enhancement factors $A'$ and $B'$ 
in front of the light modulus contribution. 
Due to these factors, the mirage messenger scale of 
mixed modulus-anomaly mediation~\cite{Choi:2005uz} for the 
visible sector SUSY breaking can reside in a wide range 
depending on the magnetic flux of $D$-brane $m_{a,b}$ and 
also on the three-form flux $G_3$ through the vacuum value 
of dilaton $S$ stabilized by the flux induced superpotential. 
On the other hand, when both $S$ and $T$ remain as light moduli, 
we have a SUSY breaking local minimum, and then the modulus 
mediated SUSY breaking is dominant after uplifting it to 
the Minkowski minimum. 

\subsection*{Acknowledgements}
The author would like to thank Tetsutaro~Higaki and Tatsuo~Kobayashi 
for the collaborations~\cite{Abe:2005rx,Abe:2005pi,Abe:2006xi} 
which form the basis of this talk, and the organizers of
SUSY06 and Summer Institute 2006. 
The author is supported by the Japan Society for the Promotion 
of Science for Young Scientists (No.0602496).


\begin{thebibliography}{99}

\bibitem{Abe:2005rx}
  H.~Abe, T.~Higaki and T.~Kobayashi,
  Phys.\ Rev.\ D {\bf 73}, 046005 (2006)
  [hep-th/0511160].

\bibitem{Abe:2005pi}
  H.~Abe, T.~Higaki and T.~Kobayashi,
  Nucl.\ Phys.\ B {\bf 742}, 187 (2006)
  [hep-th/0512232].

\bibitem{Abe:2006xi}
  H.~Abe, T.~Higaki and T.~Kobayashi,
  Phys.\ Rev.\ D {\bf 74}, 045012 (2006)
  [hep-th/0606095].

\bibitem{Lust:2004cx}
  D.~Lust, P.~Mayr, R.~Richter and S.~Stieberger,
  Nucl.\ Phys.\ B {\bf 696}, 205 (2004)
  [hep-th/0404134].

\bibitem{Gukov:1999ya}
  S.~Gukov, C.~Vafa and E.~Witten,
  Nucl.\ Phys.\ B {\bf 584}, 69 (2000)
  [Erratum-ibid.\ B {\bf 608}, 477 (2001)]
  [hep-th/9906070].

\bibitem{Giddings:2001yu}
  S.~B.~Giddings, S.~Kachru and J.~Polchinski,
  Phys.\ Rev.\ D {\bf 66}, 106006 (2002)
  [hep-th/0105097].

\bibitem{Kachru:2003aw}
  S.~Kachru, R.~Kallosh, A.~Linde and S.~P.~Trivedi,
  Phys.\ Rev.\ D {\bf 68}, 046005 (2003)
  [hep-th/0301240].

\bibitem{Choi:2005ge}
  K.~Choi, A.~Falkowski, H.~P.~Nilles and M.~Olechowski,
  Nucl.\ Phys.\ B {\bf 718}, 113 (2005)
  [hep-th/0503216].

\bibitem{Choi:2005uz}
  K.~Choi, K.~S.~Jeong and K.~i.~Okumura,
  JHEP {\bf 0509}, 039 (2005)
  [hep-ph/0504037].


\end{thebibliography}
\end{document}